\documentclass[fleqn,usenatbib]{mnras}
\usepackage[english]{babel}
\usepackage{lmodern}
\usepackage{microtype}
\usepackage{booktabs}
\usepackage{bm}
\usepackage{graphicx}
\usepackage{amsfonts}
\usepackage{amsmath}
\usepackage{amssymb}
\usepackage{listings}
\usepackage{physics}
\usepackage{subfigure}
\usepackage{fancyhdr}
\usepackage{xcolor}

\usepackage{standalone}

\title[FIGARO]{Rapid localization of gravitational wave hosts with FIGARO}

\author[S. Rinaldi and W. Del Pozzo]{
Stefano Rinaldi$^{1,2}$\thanks{E-mail: stefano.rinaldi@phd.unipi.it}
and Walter Del Pozzo$^{1,2}$
\\
$^{1}$Dipartimento di Fisica ``E. Fermi", Università di Pisa, I-56127 Pisa, Italy\\
$^{2}$INFN, Sezione di Pisa, I-56127 Pisa, Italy
}
\date{\today}

\date{Accepted XXX. Received YYY; in original form ZZZ}

\pubyear{2022}

\begin{document}
\label{firstpage}
\pagerange{\pageref{firstpage}--\pageref{lastpage}}
\maketitle
\bibliographystyle{mnras}

\begin{abstract}
    The copious scientific literature produced after the detection of GW170817 electromagnetic counterpart demonstrated the importance of a prompt and accurate localization of the gravitational wave within the co-moving volume. In this letter, we present \textsc{figaro}, a ready to use and publicly available software that relies on Bayesian non-parametrics. \textsc{figaro} is designed to run in parallel with parameter estimation algorithms to provide updated three-dimensional volume localization information. Differently from any existing algorithms, the analytical nature of the \textsc{figaro} reconstruction allows a ranking of the entries of galaxy catalogues by their probability of being the host of a gravitational wave event, hence providing an additional tool for a prompt electromagnetic follow up of gravitational waves. We illustrate the features of \textsc{figaro} on binary black holes as well as on GW170817. Finally, we demonstrate the robustness of \textsc{figaro} by producing so-called pp-plots and we present a method based on information entropy to assess when, during the parameter estimation run, it is reasonable to begin releasing skymaps.
\end{abstract}


\section{Introduction}
The discovery of the binary neutron star (BNS) merger GW170817 \citep{gw170817discovery} by the LIGO~\citep{LIGOdetector} and Virgo~\citep{VIRGOdetector} detectors and the simultaneous detection of the short gamma-ray burst (GRB) GRB 170817A \citep{grb-fermi, grb-fermilv} led to an intense follow up campaign \citep{emcampaign} that ultimately identified the kilonova transient AT2017gfo~\citep{coulter} and NGC 4993 as the host galaxy. 

This multi-messenger observation was a scientific goldmine, leading to the confirmation that BNS mergers are progenitors of short GRBs, as predicted decades earlier~\citep[e.g.][]{goodman, paczynski, eichler}, that the aftermath of the merger leads to intense r-processes~\citep{pian2017} and allowed a first independents determination of the Hubble constant~\citep{lvk_h0}.

These results, which are a selected subset of the vast literature that sparked afterwards, have been made possible by decade-long methodological and technological efforts. From the GW side, which is our area of expertise, the discovery of GW170817 was enabled by not only the detectors, but by the development of online search pipelines, such as \textsc{MBTA} \citep{Aubin:2020goo}, \textsc{pycbc} \citep{Usman:2015kfa}, \textsc{gstlal} \citep{Sachdev:2019vvd} and \textsc{cWB} \citep{Klimenko:2015ypf}, that were able to identify the GW trigger almost instantly. The trigger activated then rapid localization algorithms such as \textsc{bayestar} \citep{bayestar} to produce a first estimate of the source location to release to the astronomical community for it to begin the hunt for a counterpart. At the same time, more in-depth parameter estimation algorithms, such as LALInference \citep{LALinference} or Bilby \citep{bilby, bilby-gwtc2}, start their endeavour to update and refine the initial \textsc{bayestar} volume maps and hence ease the astronomers hunt for a counterpart \citep{singer2016, singer2016-supplement}. 

This whole process happened -- and still does happen -- on timescales that range from seconds to minutes, from the online search to the \textsc{bayestar} initial volume map, to hours for a first accurate parameter estimation release of the GW source position in the volume.
GW170817 demonstrates that the whole process is extremely successful. 
However, the hours gap in between the first \textsc{bayestar} map and the final Markov Chain Monte Carlo (MCMC) is, in our opinion, still an unacceptably long time for the astronomers community to wait for updated localization information.

In this letter, we propose a solution to this problem. Our solution is based on Bayesian non-parametrics and led to the development of an algorithm which we named \textsc{figaro} for \textsc{\emph{Fast Inference for GW Astronomy, Research \& Observations}}. \textsc{figaro} is publicly available at \url{https://github.com/sterinaldi/figaro}. 
The main features, and point of departures from any previous efforts, of \textsc{figaro} are the reconstruction of the three-dimensional posterior distribution over sky position and luminosity distance of a GW source from partial MCMC results in terms of analytical functions (a mixture of multivariate normal distributions) that can be used to (i) estimate credible regions in the three and two dimensional localization space; (ii) assess the convergence of the underlying MCMC by looking at the information entropy of the reconstructed distribution; (iii) evaluate the probability than \emph{any} galaxy is the host of the GW event without relying on a pixelisation \citep[e.g.][]{healpix} of the sky. All of the above points happen simultaneously with an MCMC run and add insignificant overhead to the sampler itself.

The rest of the letter is thus organised as follows: in Sec.~\ref{sec:densityestimate} we briefly review the Dirichlet process Gaussian mixture model that is the backbone of \textsc{figaro}. In Sec.~\ref{sec:volrec} we describe our solution to the volume localization of a GW source in real time (with respect to a MCMC run) and the generation of the probability ranked list of potential galaxy hosts. We conclude with some considerations in Sec.~\ref{sec:conclusions}. The more technical points, such as the use of the information entropy to assess the convergence of \textsc{figaro} as well as an analysis of its robustness via so-called pp-plots are presented in Supplementary Material, Appendix A and B, respectively.

\section{Density estimate}\label{sec:densityestimate}
In this section we briefly review the Dirichlet process (DP) and Dirichlet process Gaussian mixture model (DPGMM) to provide the reader with the necessary background for \textsc{figaro}. However, a comprehensive description of the model and computational method is beyond the scope of this letter, hence we refer the interested reader to~\citet{delpozzo2018}, where a similar idea was first explored, and to~\citet{hdpgmm} where the details of the method (and its hierarchical generalisation) and of the sampling algorithm are described.

To develop an intuition of a DP and of the DPGMM, begin by considering an experiment whose outcomes can assume a finite number of values, say $K$, and imagine drawing a certain number of samples $N$. Each of the $K$ possible outcomes can be thought of as a separate, known a priori, ``class'' of outcomes.
If the samples are \emph{exchangeable}, the order in which they are drawn does not matter; the only relevant information is how many times we observed each of the $K$ classes, $\mathbf{n} = \{n_1,\ldots, n_K\}$ with $\sum_i^K n_i = N$.

Given $\mathbf{n}$, we wish to infer the probabilities $\mathbf{q} = \{q_1,\ldots, q_K\}$ associated with each of the $K$ classes. This is done via Bayes' theorem:
\begin{equation}
    p(\mathbf{q}|\mathbf{n}) \propto p(\mathbf{n}|\mathbf{q})p(\mathbf{q})\,.
\end{equation}
where the likelihood for the observed counts is a multinomial distribution
\begin{equation}\label{multinomial}
    p(\mathbf{n}|\mathbf{q}) = \frac{N!}{\prod_i^Kn_i!}\prod_i^K q_i^{n_i}\,.
\end{equation}
and the prior distribution for the probabilities $\mathbf{q}$ is a Dirichlet distribution (DD)
\begin{equation}\label{dirichletdistribution}
    p(\mathbf{q}|\mathbf{h},\alpha) = \mathrm{Dir}(\mathbf{q}|\boldsymbol{\alpha}) = \frac{\Gamma(\alpha)}{\prod_i^K\Gamma(\alpha h_i)}\prod_i^K q_i^{\alpha h_i -1}\,,
\end{equation}
where $\mathbf{h} = \{h_1,\ldots,h_K\}$ is a discrete probability distribution, hence $\sum_ih_i = 1$. 
The vector $\mathbf{h}$ and the parameter $\alpha$ completely characterise the DD:  $\mathbf{h}$ is the mean -- expected value -- for $\mathbf{q}$ and $\alpha$ quantifies how concentrated around $\mathbf{h}$ the vector $\mathbf{q}$ is.
Being conjugate to the multinomial distribution, the posterior on $\mathbf{q}$ is still a DD:
\begin{equation}\label{posteriorDD}
    p(\mathbf{q}|\mathbf{n},\mathbf{h},\alpha) = \frac{\Gamma(\alpha + N)}{\prod_i^K \Gamma(\alpha h_i + n_i)}\prod_i^K q_i^{n_i+\alpha h_i-1}\,.
\end{equation}

Consider now the case is which the number of ``classes'' is unknown. Hence, a priori, they could be countably infinite. The problem can still be solved by introducing the Dirichlet process (DP), the infinite dimensional generalisation of the DD. Assigning probabilities to an infinite set might seem an hopeless task, however an observer will \emph{always} record a finite number of outcomes. This finite set of outcomes will then be described by a DD. However, the DP takes into account the possibility that there are -- a formally infinite -- set of outcomes that have not been recorded yet. Similarly to the DD, the DP is parametrized by a (continuous) base distribution $H$, which plays the role of $\mathbf{h}$, and a concentration parameter $\alpha$, as before. A DP is particularly useful in the definition of the Dirichlet process Gaussian mixture model~\citep{escobar&west}. The purpose is to represent arbitrary probability distributions as an infinite mixture of multivariate Gaussian distributions~\citep{rasmussen}:
\begin{equation}\label{gmm}
    p(\mathbf{x}) = \sum_{k=1}^{\infty} w_k \mathcal{N}(\mathbf{x}|\boldsymbol\mu_k,\boldsymbol\sigma_k)\,,
\end{equation}
where $\sum_k w_k = 1$ and $\boldsymbol\mu_k$ and $\boldsymbol\sigma_k$ are the mean vector and covariance matrix of a multivariate Gaussian distribution, denoted by $\mathcal{N}$. The way in which the DP enters in the picture is by providing a practical way of inferring the mixing proportions \emph{and} relative means $\boldsymbol\mu = \{\boldsymbol\mu_k\}_{k=1,\ldots,\infty}$ and covariances $\boldsymbol\sigma = \{\boldsymbol\sigma_k\}_{k=1,\ldots,\infty}$ from some observation $\mathbf{X}$. Applying Bayes' theorem:
\begin{equation}\label{pofprob}
    p(\mathbf{w},\boldsymbol\mu,\boldsymbol\sigma|\mathbf{X}) \propto p(\mathbf{X} |\mathbf{w},\boldsymbol\mu,\boldsymbol\sigma) p(\mathbf{w},\boldsymbol\mu,\boldsymbol\sigma) \,.
\end{equation} 

Drawing samples from a Gaussian mixture is done via Gibbs sampling: draw one of the components with probability $w_k$ and from that component draw a sample from the Gaussian distribution. Each sample will be therefore associated with a specific component. This information is stored in an indicator variable $z_i$ containing the label of the component the sample has been drawn from. Given a set of samples $\mathbf{X}$, the vector $\mathbf{z} = \{z_1,\ldots,z_N\}$ and assuming a symmetric -- uniform in the $\mathbf{w}$ -- DP prior on $\mathbf{w}$, it is possible to write the probability distribution for $\mathbf{w}$,
\begin{equation}\label{wposterior}
    p(\mathbf{w}|\mathbf{z}, \alpha) = \frac{\Gamma(N+\alpha)}{\prod_i^K\Gamma(n_i+\alpha/N)}\prod_i^Kw_i^{n_i+\alpha/N-1}\,,
\end{equation}
where $n_i$ denotes the number of samples associated with component $i$, and for $\boldsymbol\mu_i$ and $\boldsymbol\sigma_i$:
\begin{equation}\label{musigmaposterior}
    p(\boldsymbol\mu_i,\boldsymbol\sigma_i|\mathbf{X},\mathbf{z}) \propto p(\{\mathbf{x}_j|z_j = i\}|\boldsymbol\mu_i,\boldsymbol\sigma_i,\mathbf{z})p(\boldsymbol\mu_i,\boldsymbol\sigma_i)\,.
\end{equation}

It is also possible to derive the conditional probability that a new data point $\mathbf{x}_{N+1}$ is drawn from one of the already observed mixture components
\begin{subequations}\label{p_z}
\begin{multline}\label{pobs}
p(z_{N+1} = j|\mathbf{x}_{N+1},\mathbf{z},\mathbf{X},\alpha) =\\= \frac{n_j}{N+\alpha}\int p(\mathbf{x}|\boldsymbol\mu_j,\boldsymbol\sigma_j,\mathbf{X},\mathbf{z})p(\boldsymbol\mu_j,\boldsymbol\sigma_j|\mathbf{X},\mathbf{z})\dd\boldsymbol\mu_j\dd\boldsymbol\sigma_j
\end{multline}
as well as from a previously unobserved component
\begin{equation}\label{punobs}
p(z_{N+1} = \mathrm{new}|\mathbf{x}_{N+1},\mathbf{z},\mathbf{X},\alpha) = \frac{\alpha}{N+\alpha}\,.
\end{equation}
\end{subequations}
The practical power of a DPGMM is this: there is no need to assign a priori an expected number of components to explain the observed data, but we can let the data tell us how many we need.
Therefore, to get a draw from~\eqref{pofprob} given a set of samples $\mathbf{X}$ from the target distribution $p(\mathbf{x})$:
\begin{enumerate}
    \item Draw a sample of $\mathbf{z}$: starting with no samples at all, iteratively add one sample at a time to the mixture, drawing $z_i$ from Eqs.~\eqref{p_z} conditioned only on the previously added samples;
    \item Once all the samples have been added, draw a sample of $\mathbf{w}$, $\boldsymbol\mu$ and $\boldsymbol\sigma$ from Eqs.~\eqref{wposterior} and~\eqref{musigmaposterior} conditioned on the previously drawn $\mathbf{z}$ or take the expected value from the same distributions\footnote{We opted for the latter, in our implementation of this algorithm.};
    \item Repeat these two steps, always starting from an empty mixture and shuffling the order in which the samples are added to it, until the desired number of $\mathbf{w}$, $\boldsymbol\mu$ and $\boldsymbol\sigma$ is accumulated.
\end{enumerate}
Since the probability distributions for $\mathbf{w}$, $\boldsymbol\mu$ and $\boldsymbol\sigma$ are conditioned on the observed $\mathbf{z}$, they are updated every time a new sample is added to the pool. Hence, if the samples are drawn by some sampling algorithm -- a MCMC algorithm, for instance, or any kind of sampling algorithm that is able to produce independent samples during its run -- a DPGMM can be updated \emph{as the sampler is running}, therefore we can estimate the posterior distribution at any stage of the run itself to provide up-to-date estimates of credible regions, as well as an analytical representation of the yet unknown MCMC target distribution. 

In a nutshell, this is what \textsc{figaro} does: by reading the current state of any MCMC sampler, it provides a DPGMM representation of the posterior distribution for the sky position and luminosity distance to a GW source \emph{at the stage at which the sampler is}. 
It is therefore possible to produce intermediate volume maps and update the localization information on a short time scale than having to wait for the expensive MCMC runs to finish.

\section{Volume reconstruction and sky localization}\label{sec:volrec}
While in principle \textsc{figaro} could be applied to the full parameter space, for this work we focus on the sky position (right ascension $\alpha$ and declination $\delta$) and luminosity distance $D_L$. Our algorithm proceeds as follows; as a MCMC is running, \textsc{figaro} reads the current set of samples from the posterior for $\alpha,\delta$ and $D_L$ and, using the density estimation scheme we described in the previous section, it reconstructs the current snapshot of the full three dimensional posterior distribution as a mixture of multivariate Gaussian distributions. \textsc{figaro} then proceeds at evaluating the 3D map on a user-defined grid that can be marginalised to obtain a two-dimensional sky map for the location of the GW host. The procedure is repeated at any fixed number of new MCMC samples, until we estimate the convergence of the density estimation via the information entropy, see Supplementary Material, Appendix A. We decided to work in terms of number of MCMC samples since the actual wall time depends quite dramatically on the waveform model adopted for the analysis as well as on the sampler used, \citep[e.g.][Table VI]{LALinference}. In the most optimistic case, one can adopt a wall time per independent sample of 3 s to get an idea of the actual time. The computational time added by \textsc{figaro} is negligible being, for a 3-dimensional distribution, approximately $500\ \mathrm{\mu s}$ per sample.

It is important, however, to point out the fact that the samples from a MCMC must be handled with care. The initial part of a chain, the so-called \emph{burn-in}, is not representative of the true distribution: the information entropy of the distribution (see appendix A), however, heavily fluctuates during the burn-in phase, indicating that the map is still unreliable. A second, more important, issue to address is the correlation among samples produced by the MCMC -- one of the founding assumption of the DPGMM is the fact that the samples are independent. We recommend to ensure, e.g. by mean of some autocorrelation computation script, that the samples fed to \textsc{figaro} are independent. However, we observed that even in the case of correlated samples, the main features of the target distribution, such as 90\% credible regions, are approximately correct.

\begin{figure*}
    \begin{subfigure}
    \centering
    \includegraphics[width = 0.6\columnwidth]{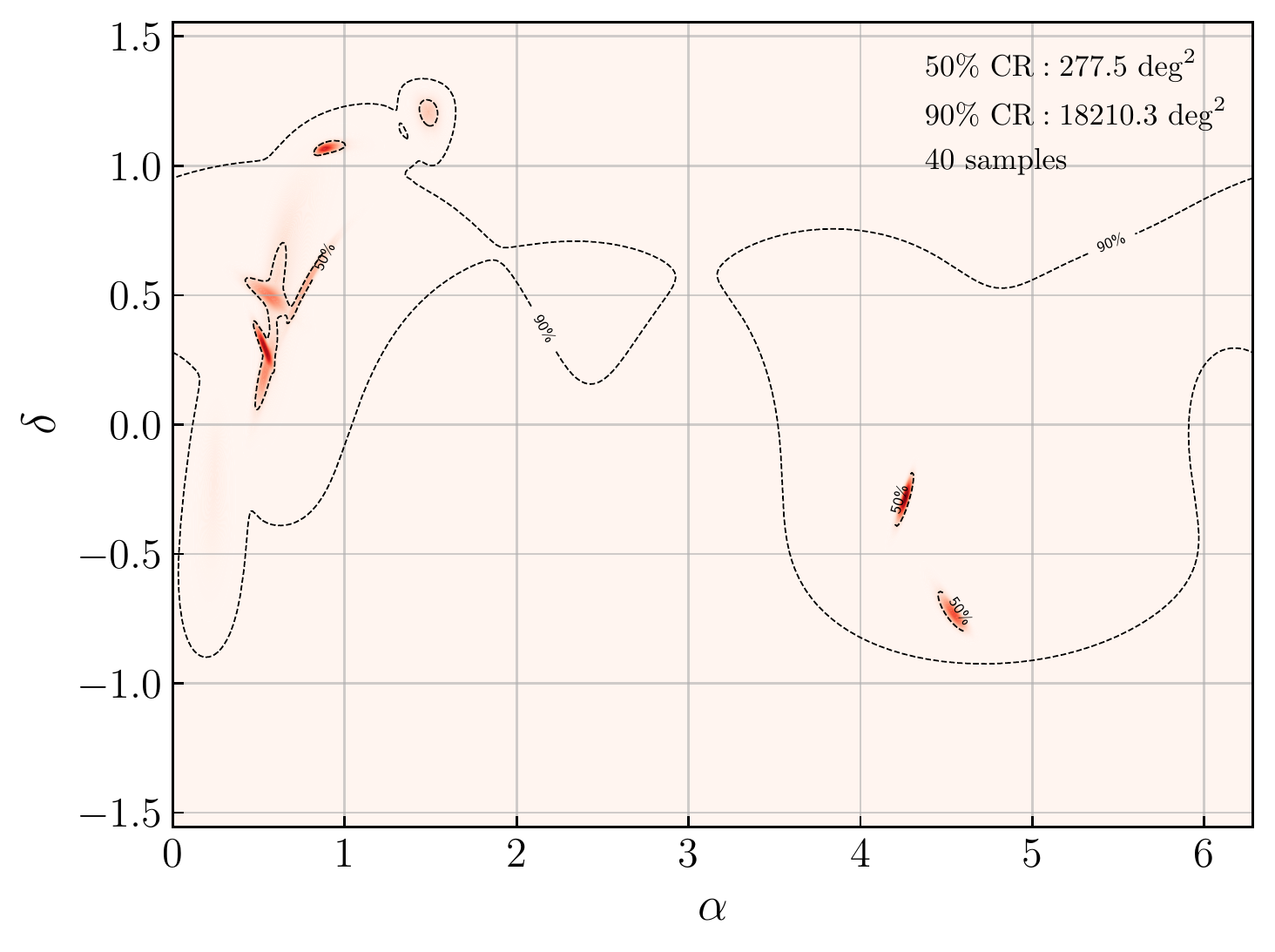}
    \end{subfigure}
    \begin{subfigure}
    \centering
    \includegraphics[width = 0.6\columnwidth]{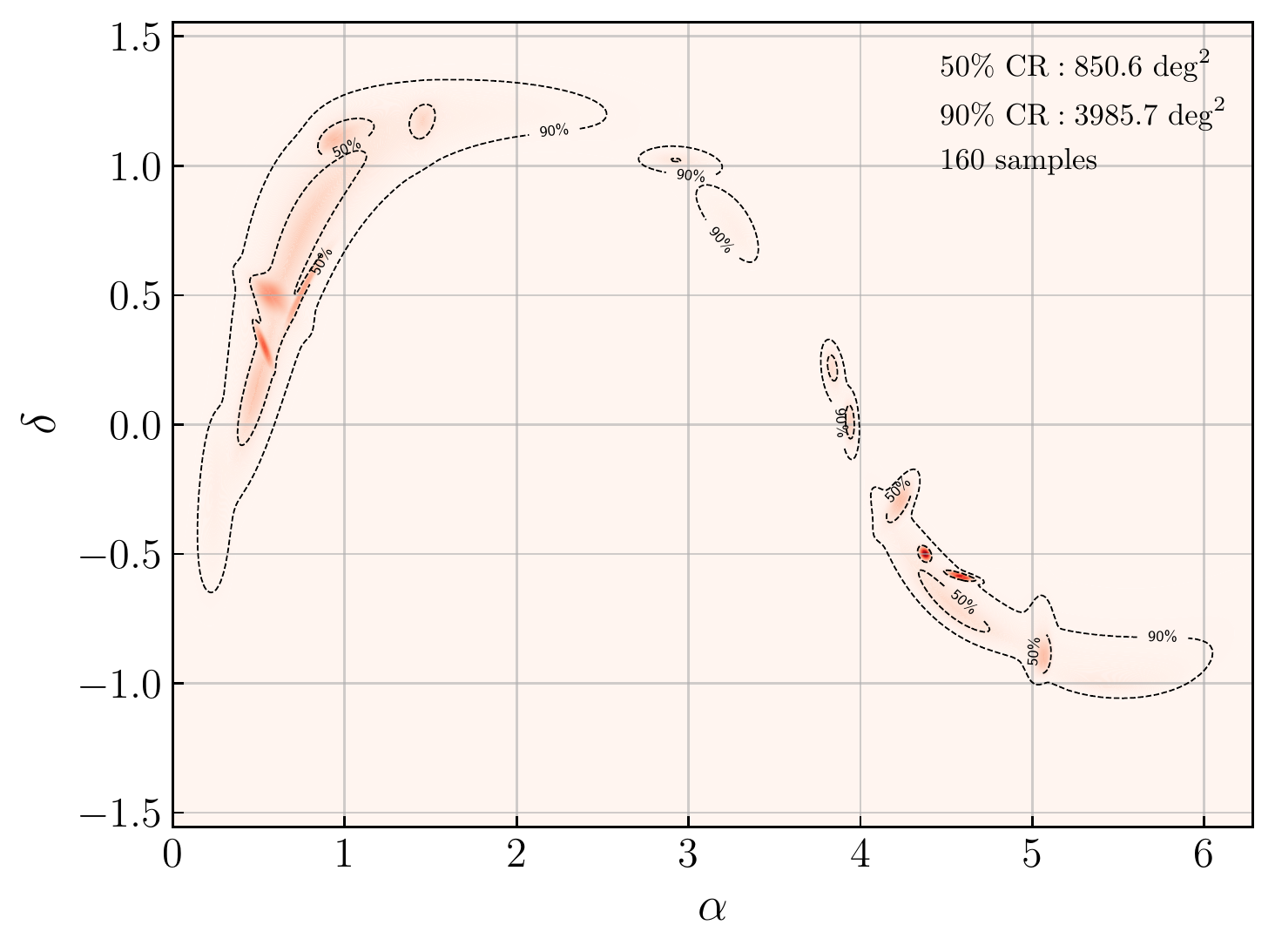}
    \end{subfigure}
    \begin{subfigure}
    \centering
    \includegraphics[width = 0.6\columnwidth]{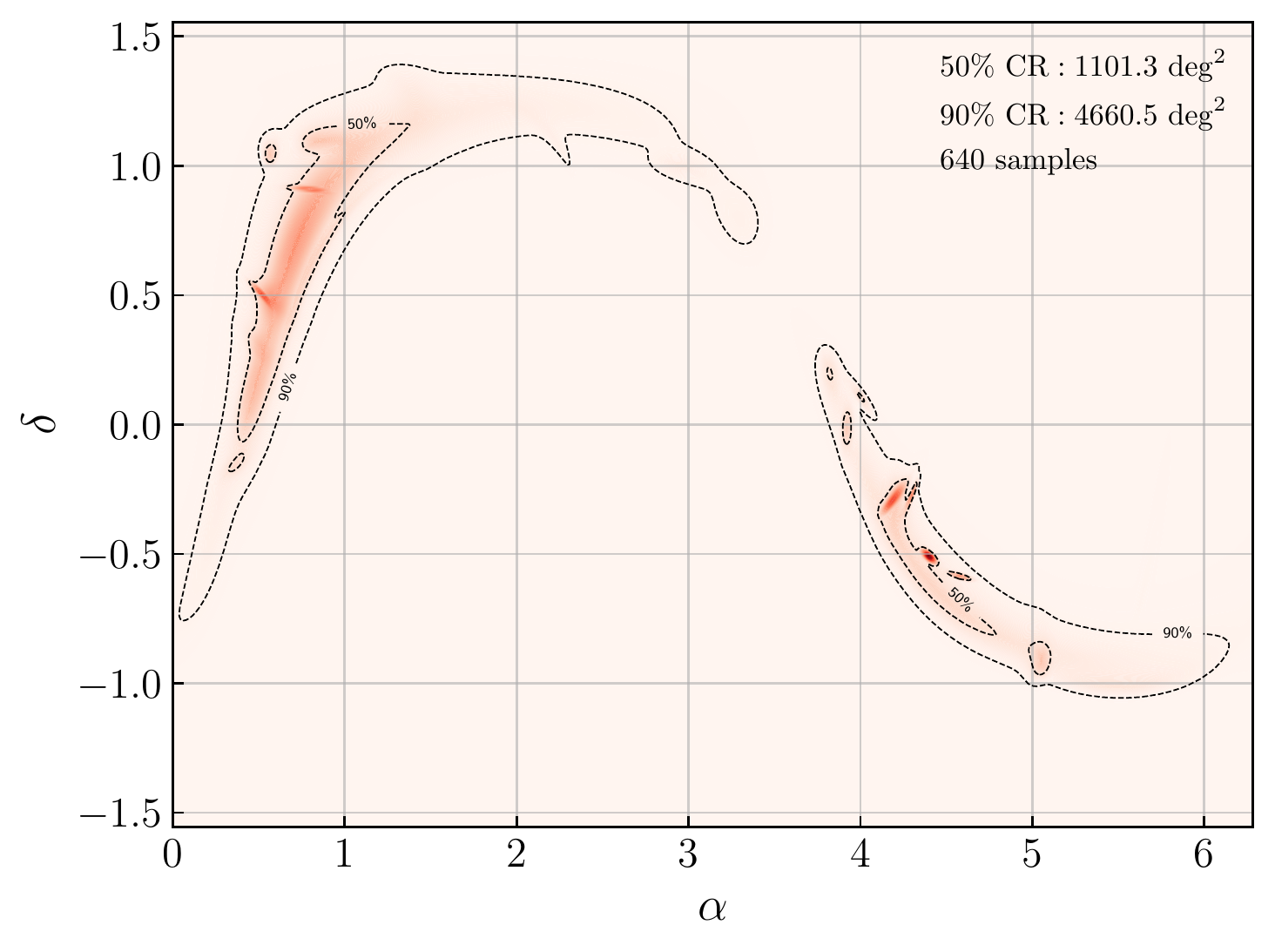}
    \end{subfigure}
    \begin{subfigure}
    \centering
    \includegraphics[width = 0.6\columnwidth]{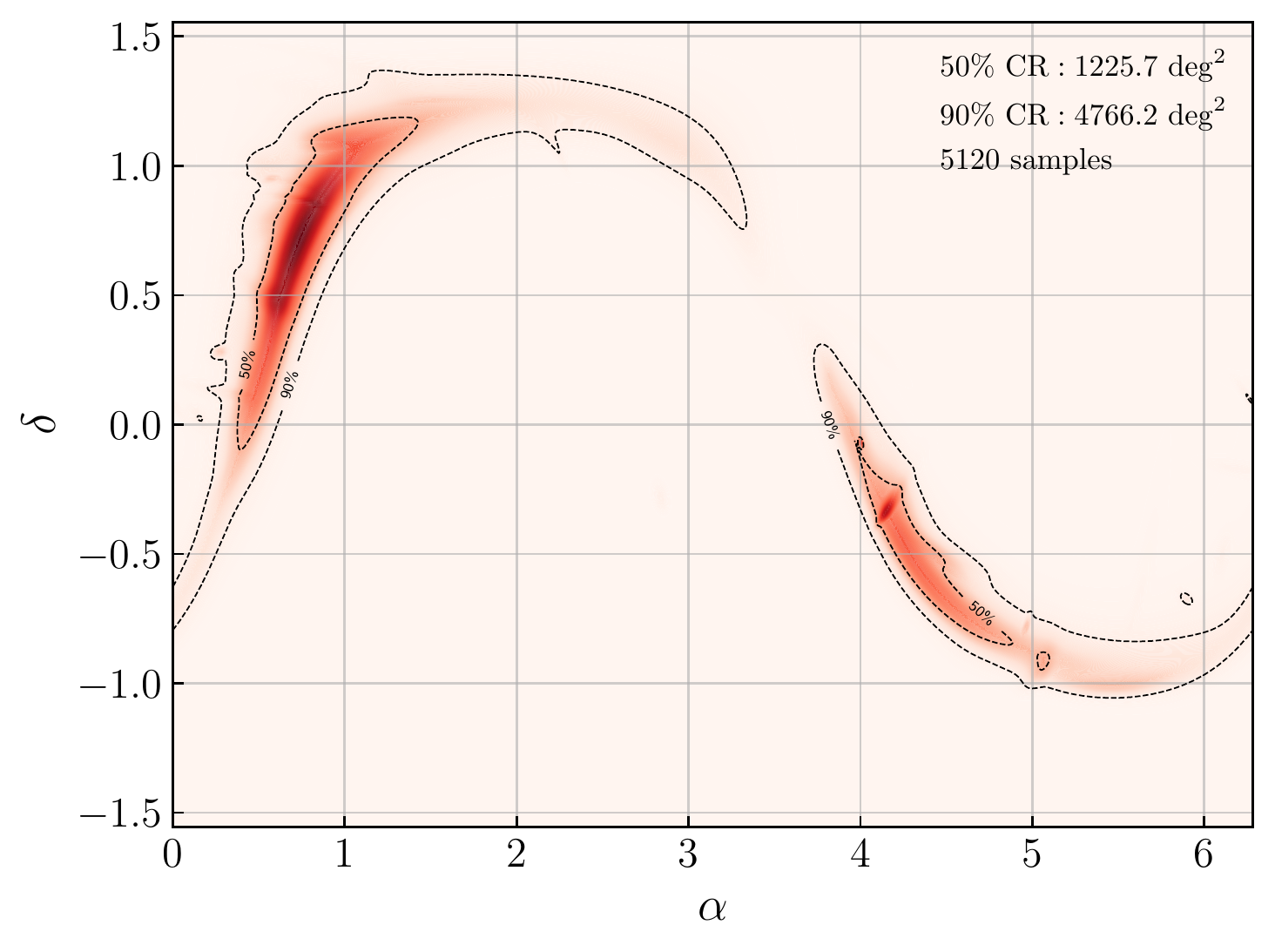}
    \end{subfigure}
    \begin{subfigure}
    \centering
    \includegraphics[width = 0.6\columnwidth]{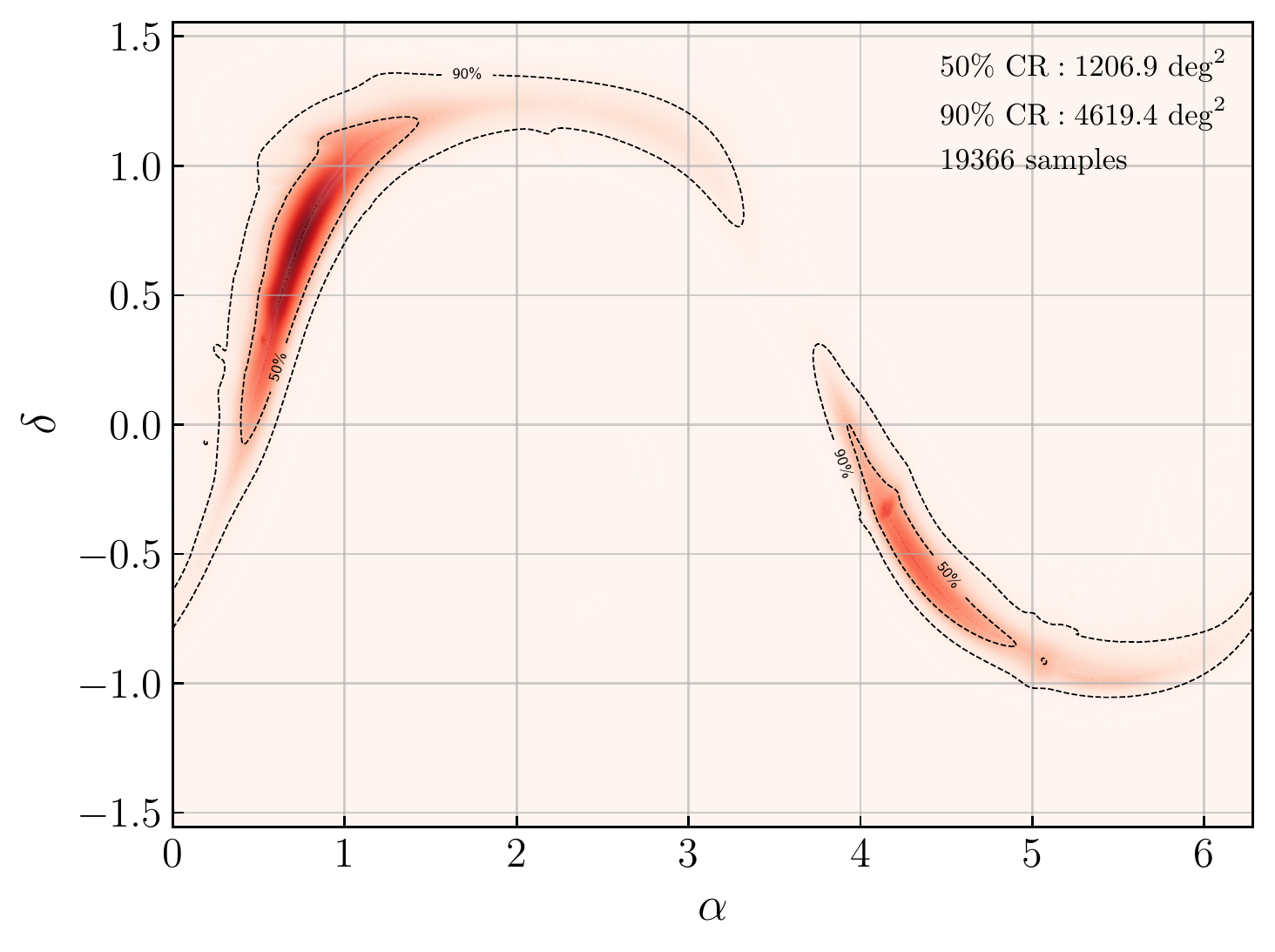}
    \end{subfigure}
\caption{Evolution of GW190909 skymap with the number of MCMC samples. The number of available samples goes from 40 to about 8000 (indicated in each panel) and increases from left to right. As the number of samples included in the analysis increases, the credible regions reconstructed by \textsc{figaro} evolve accordingly.\label{fig:190909}}
\end{figure*}

\begin{figure}
    \centering
    \includegraphics[width = \columnwidth]{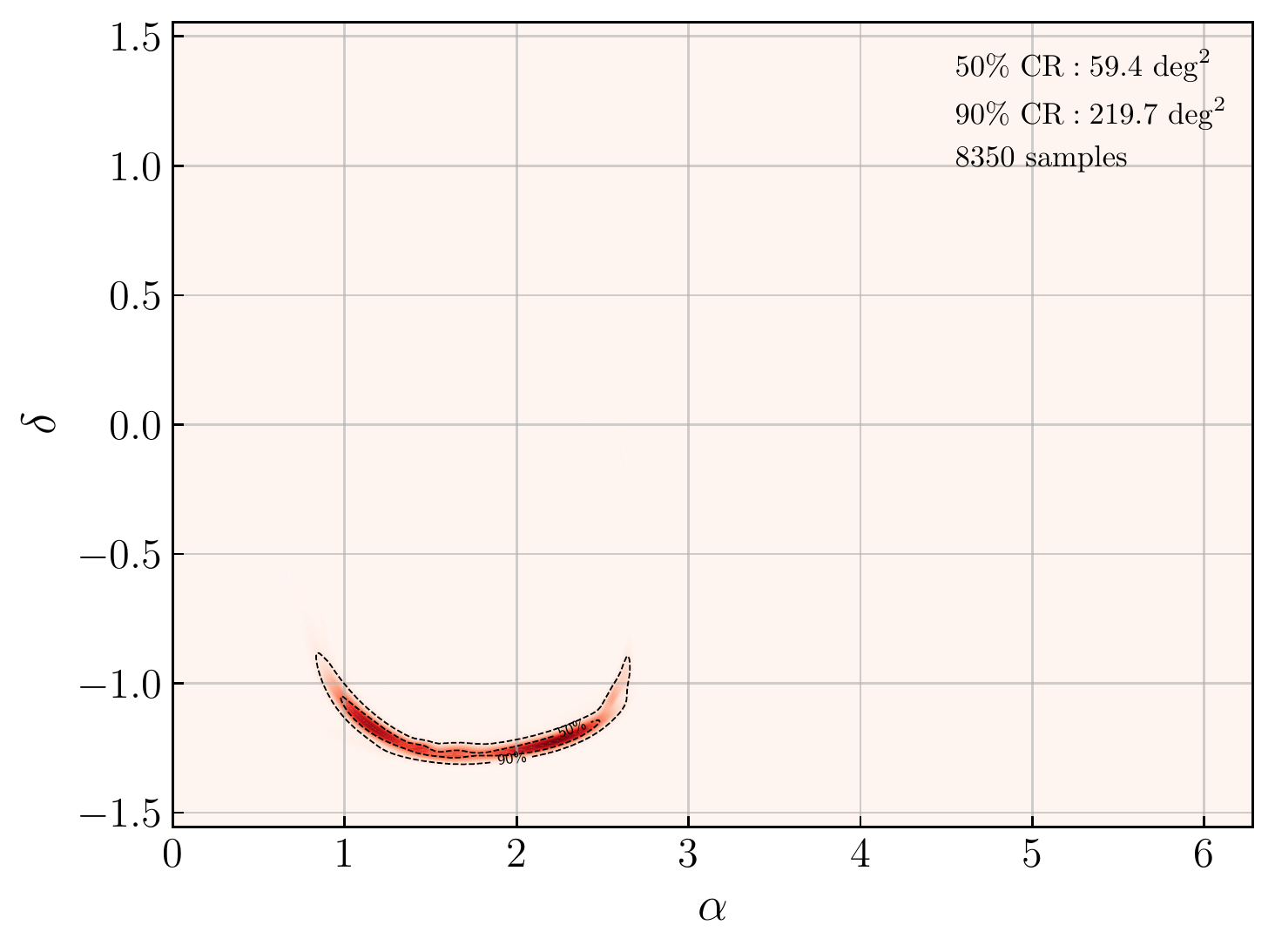}
    \caption{Skymap for GW150914. We obtain a 90 per cent credible region of 219.7 $\mathrm{deg}^2$, whereas \citet{gwtc1} reports, for the same event, a credible region of 182 $\mathrm{deg}^2$.\label{fig:150914}}
\end{figure}

To demonstrate the efficacy of our proposed scheme, we analysed the publicly available posterior samples released on GWOSC \citep{gwosc} and simulated the above approach by producing results from \textsc{figaro} using an incremental number of samples. An example of the marginal sky position posterior for GW190909 can be seen in Fig.~\ref{fig:190909}, where we show the evolution of the 90 per cent and 50 per cent credible regions as a function of the number of samples accumulated by the MCMC. We obtain a 90 per cent credible region of 4619.4 $\mathrm{deg}^2$ whereas \citet{gwtc2} reports, for the same event, a 90 per cent credible region of 4700 $\mathrm{deg}^2$. As a further comparison, Figure~\ref{fig:150914} shows the \textsc{figaro} skymap for GW150914 \citep{gw150914} as reconstructed using the posterior samples from the gravitational wave catalogue GWTC--3 \citep{gwtc3}.

\subsection{Exploiting galaxy catalogues}\label{sec:galcat}
Current strategies for EM follow up, see for instance \citet{coughlin:2018} and references therein, imply the adoption of tiling algorithms, aimed at scheduling telescope observations in order to optimise some pre-defined metric -- e.g. minimise the number of pointings or maximise the probability of observing a given counterpart at fixed magnitude -- given the time allocation on each telescope.
As pointed out earlier, \textsc{figaro} three-dimensional density estimates are \emph{analytical}, hence they can be evaluated efficiently over any point in the parameter space to provide the value of the probability density at that point, including the standard pixelisation schemes required by tiling algorithms.

We would like to highlight that, since search optimization pipelines such as \textsc{gwemopt}, presented in \citet{coughlin:2018}, are triggered by skymaps provided by the GW localisation algorithms, as long as the format in which these skymaps are provided is homogeneous, the way in which they are produced should not affect these pipelines. Looking at the flowchart in Figure 1 of the aforementioned paper, \textsc{figaro} is positioned upstream of the GraceDB branch.

We propose to exploit the analytical form of the DPGMM posterior in order to maximise the probability of identifying the potential host of a GW event, as we shall see in this section. The rapid decay of the faint light curve for GW170817 \citep{gw170817lightcurve} and the expected similar behaviour for other kilonovae signals \citep{kilonovadetectability} suggest that a successful EM follow up campaign should plan to identify the host of the GW event as soon as possible, ideally within the first 36 hours of the GW transient. Considering that the expected detection range for the LVK network in O4 and O5 is expected to reach O(100) Mpc for BNS, the rapid host identification becomes even more critical, given the limited sensitivity of rapid EM follow up telescopes. 

Call $f(\mathbf{x})$ the DPGMM inferred by \textsc{figaro}, with $\mathbf{x}\equiv(\alpha,\delta,D_L)$; given a list of known galaxy positions $\mathbf{x}_g$, we can compute the probability densities $f(\mathbf{x}_1),\ldots,f(\mathbf{x}_n)$ and sort them in decreasing order to provide a probability-ranked list of potential hosts to a GW. The possibility of doing so in real time is a major departure from any other algorithms. Moreover, the list can be updated at any time \textsc{figaro} releases a new density estimate from an on-going MCMC run. We applied this method to GW170817 using the \textsc{GLADE+} catalog \citep{glade+}. 

\begin{figure}

    \begin{subfigure}
    \centering
    \includegraphics[width = 0.72\columnwidth]{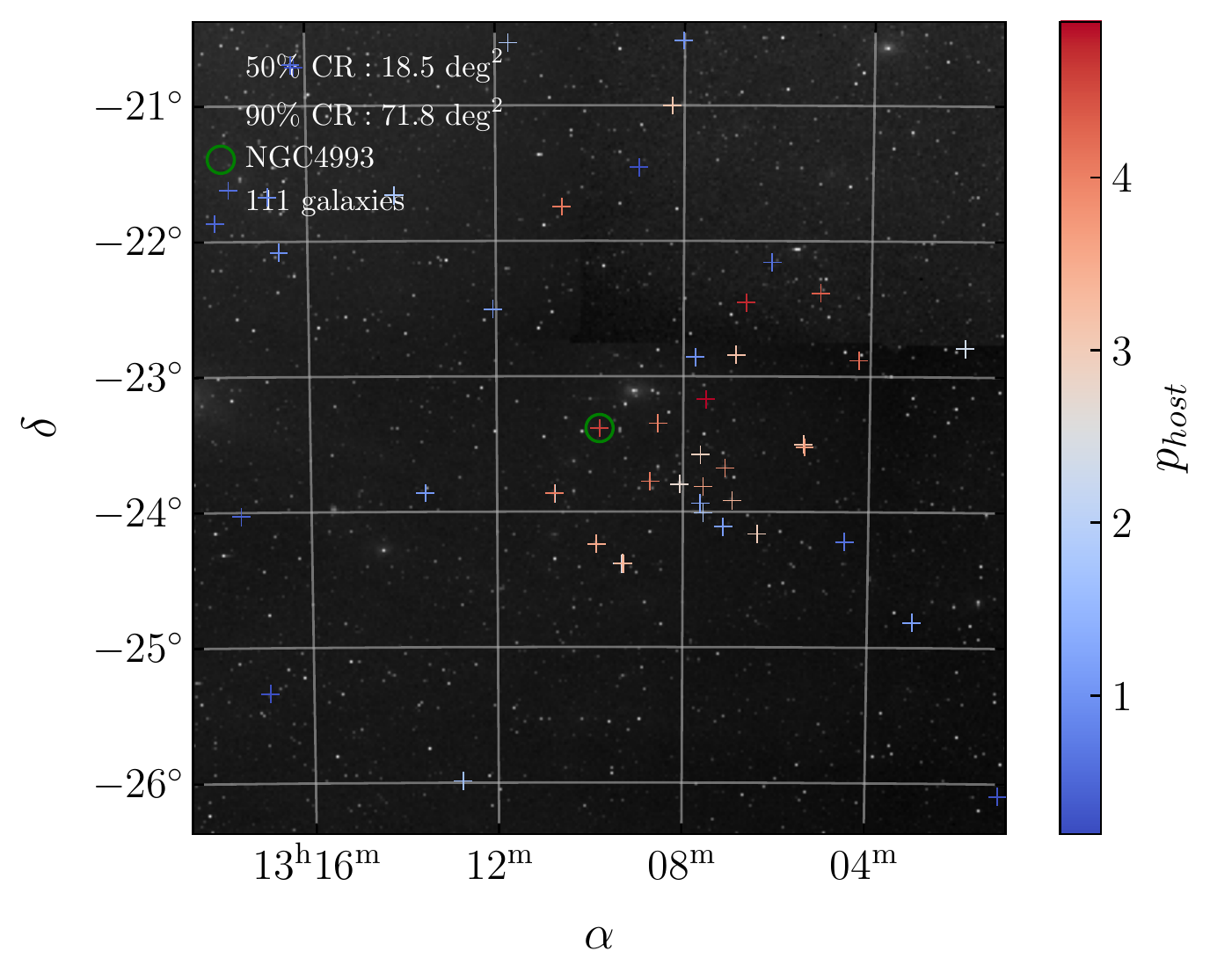}
    \end{subfigure}

    \begin{subfigure}
    \centering
    \includegraphics[width = 0.72\columnwidth]{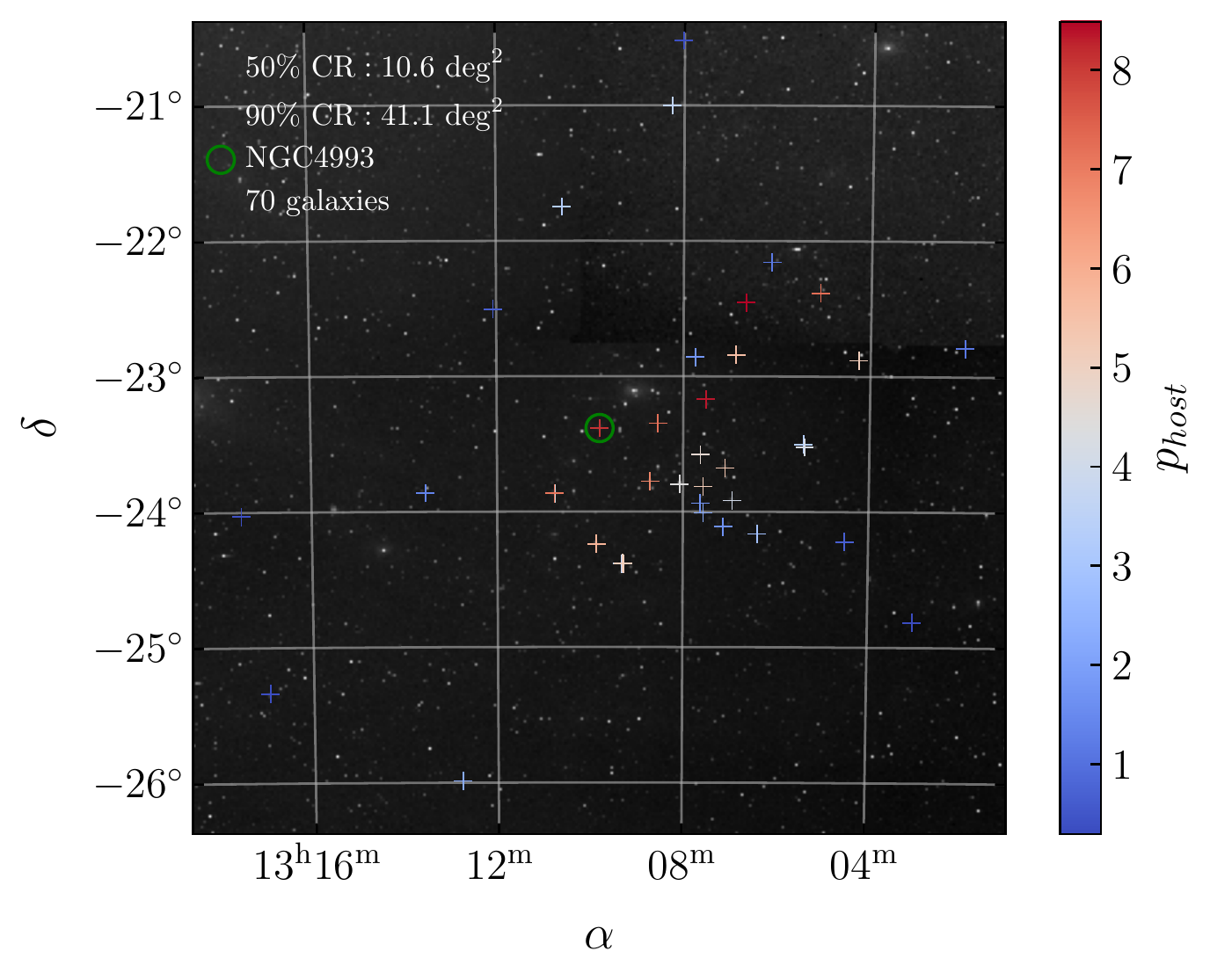}
    \end{subfigure}

    \begin{subfigure}
    \centering
    \includegraphics[width = 0.72\columnwidth]{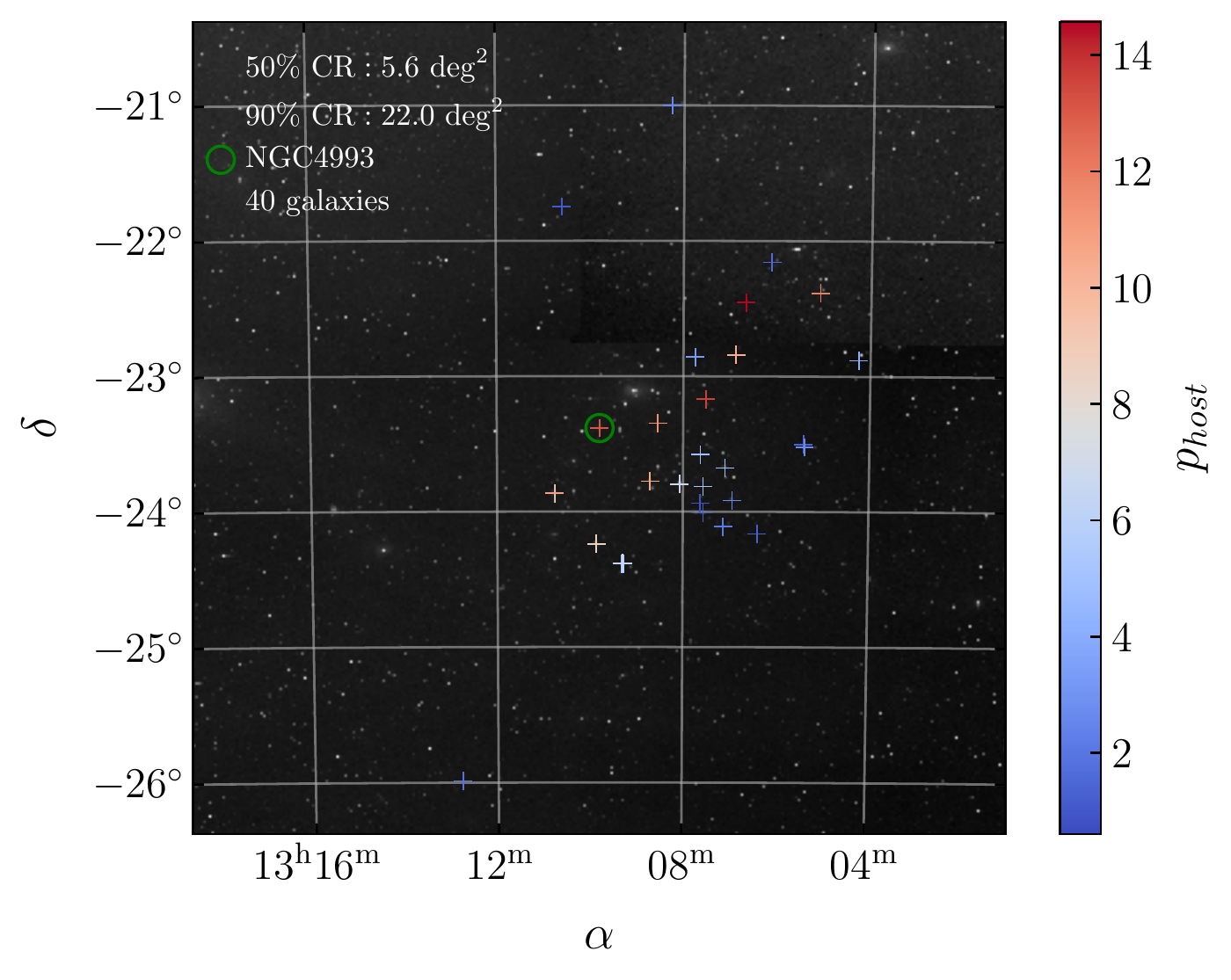}
    \end{subfigure}

    \begin{subfigure}
    \centering
    \includegraphics[width = 0.72\columnwidth]{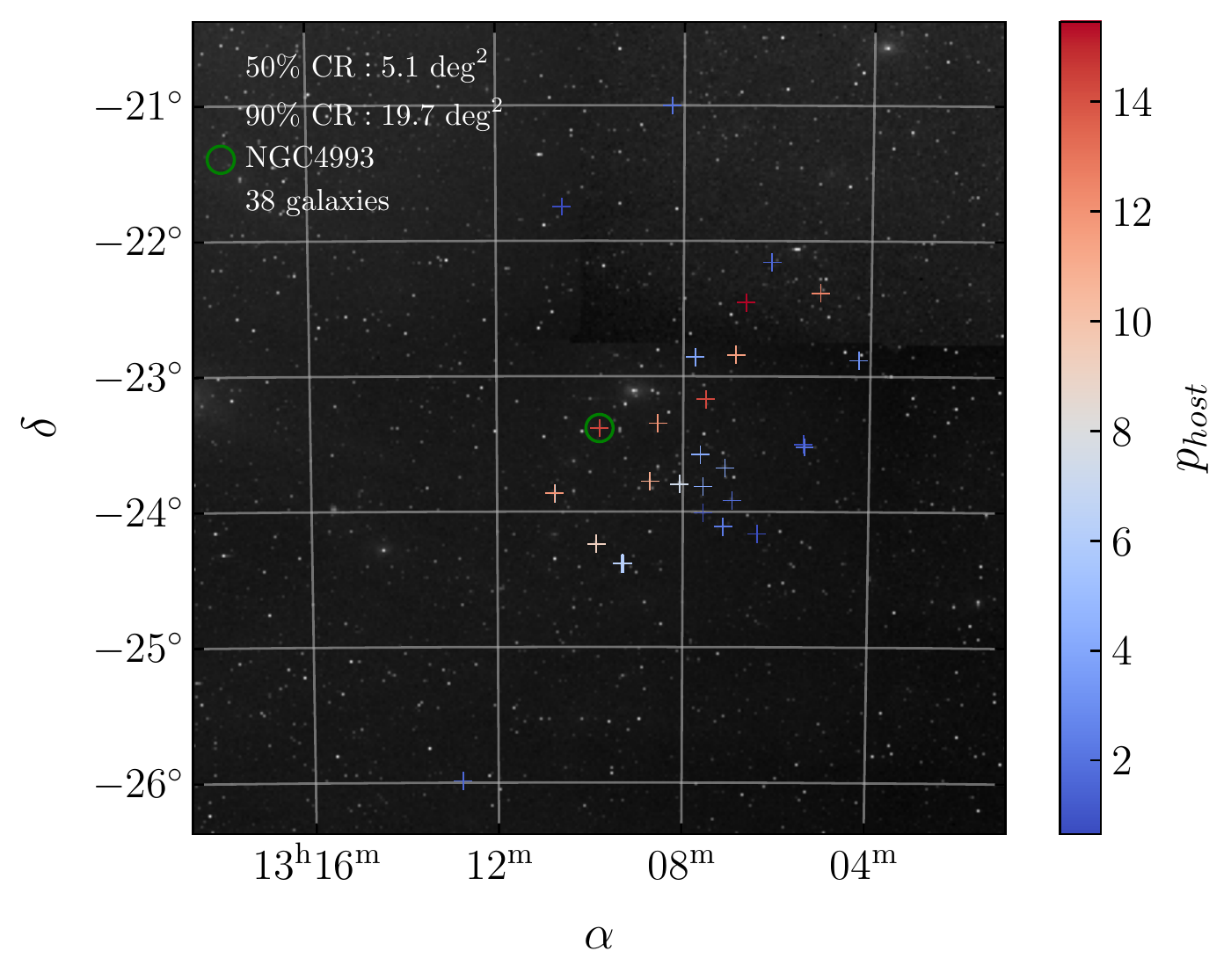}
    \end{subfigure}
\caption{Evolution of potential galaxy hosts for GW170817 with the number of PE samples. The number of samples is respectively, from top to bottom, 40, 160, 1280 and 5042. Galaxies within the 90 per cent credible volume are marked by crosses coloured according to the probability of the galaxy of being the host conditioned only on the information coming from the GW. NGC 4993 is highlighted by a green circle. The five most probable galaxies for each skymap are reported in Table~\ref{tab:catalog_170817}. Background images are taken from the Digitized Sky Surveys (DSS).\label{fig:galrank}}
\end{figure}

We followed the procedure outlined in the previous section, by progressively including more and more MCMC samples to our DPGMM inference and produced a list of potential galactic hosts together with the two and three-dimensional credible regions. The result of our experiment is shown in Fig.~\ref{fig:galrank}. This shows the evolution of the galaxies within the 90 per cent credible volume for GW170817, coloured according to the probability of being the host.
NGC4993, the host galaxy, is highlighted with a green circle.

\begin{table*}
    \centering
    \begin{tabular}{c|c|c|c}
    \toprule
    \textbf{40 samples} & \textbf{160 samples} & \textbf{1280 samples} & \textbf{5042 samples}\\
    \midrule
        LEDA 803966 & ESO 575-55 & ESO 575-55 & ESO 575-55\\
        ESO 575-55 & LEDA 803966 & LEDA 803966 & \textcolor{red}{NGC 4993}\\
        \textcolor{red}{NGC 4993} & \textcolor{red}{NGC 4993} & \textcolor{red}{NGC 4993} & LEDA 803966\\
        ESO 575-53 & ESO 575-53 & ESO 575-53 & ESO 575-53\\
        LEDA 169663 & 6dFGS gJ130832.3-232050 & 6dFGS gJ130832.3-232050 & 6dFGS gJ130832.3-232050\\
    \bottomrule
    \end{tabular}
    \caption{List of the top 5 most probable hosts to GW170817 as identified by evaluating the \textsc{figaro} density reconstruction over the GLADE+ catalogue. The true host NGC 4993 is always in the list, also extremely early in the run and it is the second most probable host when 5042 samples have been accumulated.}
    \label{tab:catalog_170817}
\end{table*}
Being GW170817 a very well localized event, the number of potential hosts for this event is quite small, some tens of galaxies. The true host turns out to be within these galaxies even with a small number of samples. We report the 5 most probable galaxies for each of the four skymaps presented in Figure~\ref{fig:galrank} in Table~\ref{tab:catalog_170817}. The true host (NGC 4993) is always in the list of the top 5 most probable hosts, even after having only 40 MCMC samples. Assuming a reasonable wall time per sample of 5s, this implies that with \textsc{figaro} after just 200s of MCMC run-time, the true host would have been the third galaxy to point at for a successful identification. 

\section{Discussion and conclusions}\label{sec:conclusions}
We presented \textsc{figaro}, a fast and reliable volume reconstruction method based on a DPGMM. We demonstrated the efficacy of the method in correctly reconstructing the two and three dimensional probability density functions and it is able to consistently estimate credible regions, as shown in Supplementary Material, Appendix B, even when declaring early convergence based on the information entropy. Furthermore, we have shown what we regard as the main advantage of \textsc{figaro} compared to other rapid localization algorithms such as the previously mentioned \textsc{bayestar}: using the analytical form of the probability density function, we can evaluate it on any available galaxy within the reconstructed volume and provide a list of potential hosts ranked by probability based on the information encoded in the GW signal. In this letter we focused exclusively on localization information from GW alone and disregarded any other information about galaxy luminosities or class. This information is, however, easily folded in the host probability calculation, e.g. by assuming an additional weight based on a Schecther luminosity function. 

This latter point also helps address one of the possible issues with our proposed ranking model for O4, O5 and subsequent observation runs. As the global network of GW detectors expands with the inclusion of KAGRA \citep{kagra} and the foreseen LIGO-India \citep{ligo-india} and its sensitivity increases, the typical distance at which potential EM counterparts also increases \citep{loctransient, pankow} and we end up in regions where the current galaxy catalogues suffer from being incomplete. This is a potential problem, although independent of our algorithm, but we do not regard this as a show stopper. It is in fact plausible to assume that if a GW is localized in a galaxy, the probability that any galaxy will be the host must be proportional to the galaxy stellar mass, hence to the galaxy luminosity~\citep{artale-1, artale-2}. In other words, more luminous galaxies are more likely to be the hosts of a GW. More luminous galaxies are also seen up to greater distances for a fixed survey sensitivity, hence are also more likely to be included in galaxy catalogues: this strategy is inquired in \citet{gehrels}. We therefore conclude that providing a list of potential hosts based on probability will still be a very valuable tool for astronomers, even when the baseline catalogue is deemed incomplete at the GW event distance.

\section*{Data availability}
\textsc{figaro} is available at \url{https://github.com/sterinaldi/figaro}. Examples and instructions on how to use the code are available on the repository.
GWTC--3 data are available in GWOSC at  \url{https://www.gw-openscience.org/}. The F2Y dataset is available at \url{https://www.ligo.org/scientists/first2years/}.

\section*{Acknowledgments}
The authors would like to acknowledge the discussions and interactions with the Parameter Estimation group and the Low-Latency group of the LIGO-Virgo-Kagra Collaboration that improved this work. We would also like to thank Christopher~P.~L.~Berry, Marica Branchesi, Edoardo Milotti, Veronica Roccatagliata, Steven~N.~Shore, Leo~P.~Singer and the anonymous referee for useful comments and discussions.

This research has made use of data or software obtained from the Gravitational Wave Open Science Center (gw-openscience.org), a service of LIGO Laboratory, the LIGO Scientific Collaboration, the Virgo Collaboration, and KAGRA. LIGO Laboratory and Advanced LIGO are funded by the United States National Science Foundation (NSF) as well as the Science and Technology Facilities Council (STFC) of the United Kingdom, the Max-Planck-Society (MPS), and the State of Niedersachsen/Germany for support of the construction of Advanced LIGO and construction and operation of the GEO600 detector. Additional support for Advanced LIGO was provided by the Australian Research Council. Virgo is funded, through the European Gravitational Observatory (EGO), by the French Centre National de Recherche Scientifique (CNRS), the Italian Istituto Nazionale di Fisica Nucleare (INFN) and the Dutch Nikhef, with contributions by institutions from Belgium, Germany, Greece, Hungary, Ireland, Japan, Monaco, Poland, Portugal, Spain. The construction and operation of KAGRA are funded by Ministry of Education, Culture, Sports, Science and Technology (MEXT), and Japan Society for the Promotion of Science (JSPS), National Research Foundation (NRF) and Ministry of Science and ICT (MSIT) in Korea, Academia Sinica (AS) and the Ministry of Science and Technology (MoST) in Taiwan.

The Digitized Sky Surveys were produced at the Space Telescope Science Institute under U.S. Government grant NAG W-2166. The images of these surveys are based on photographic data obtained using the Oschin Schmidt Telescope on Palomar Mountain and the UK Schmidt Telescope. The plates were processed into the present compressed digital form with the permission of these institutions.
The National Geographic Society - Palomar Observatory Sky Atlas (POSS-I) was made by the California Institute of Technology with grants from the National Geographic Society.
The Second Palomar Observatory Sky Survey (POSS-II) was made by the California Institute of Technology with funds from the National Science Foundation, the National Geographic Society, the Sloan Foundation, the Samuel Oschin Foundation, and the Eastman Kodak Corporation.
The Oschin Schmidt Telescope is operated by the California Institute of Technology and Palomar Observatory.
The UK Schmidt Telescope was operated by the Royal Observatory Edinburgh, with funding from the UK Science and Engineering Research Council (later the UK Particle Physics and Astronomy Research Council), until 1988 June, and thereafter by the Anglo-Australian Observatory. The blue plates of the southern Sky Atlas and its Equatorial Extension (together known as the SERC-J), as well as the Equatorial Red (ER), and the Second Epoch [red] Survey (SES) were all taken with the UK Schmidt.
Supplemental funding for sky-survey work at the ST ScI is provided by the European Southern Observatory.
\bibliography{bibliography.bib}
\bsp
\label{lastpage}
\end{document}


\label{firstpage}
\pagerange{\pageref{firstpage}--\pageref{lastpage}}
\maketitle
\bibliographystyle{mnras}

\appendix
\section{Entropy}\label{app:entropy}
Figure 1 demonstrates that, as the inference proceeds, the skymaps drastically alter their appearance during a MCMC run. We argued that to maximise the probability of detecting an EM counterpart to a GW, it is imperative to release skymaps and lists of potential targets as soon as possible. On the other hand, releasing these products too early in the run, when we know that the MCMC is still far from convergence, would be meaningless and result in wasted resources and efforts from the astronomers side.
It is therefore necessary to establish a criterion to determine the ``goodness'' of a localization, without knowing what the final posterior will be. Necessarily, we will have to define some arbitrary threshold, so there are no correct answers, but only useful ones. 
Here we propose a solution based on the information entropy content of the reconstructed distribution. The information entropy of a discrete distribution $\{p_k\}$, introduced by Shannon \citep{shannon}, is 
\begin{equation}\label{entropy}
    S = -\sum_k p_k\log{p_k}\, .
\end{equation}
and describes how much information is stored in the probability distribution itself\footnote{For the readers less familiar with information theory, we provide an intuitive interpretation of the term ``information'' in this context. The information in a probability distribution, measured typically in bits, can be thought of as the length of the message I need to send to someone, for that someone to be able to predict, in a statistical sense, the outcomes of successive draws from the given probability distribution.}. 
\textsc{figaro} uses the information change in the reconstructed probability distribution by adding every new MCMC sample to the list of observed outcomes.
As the \textsc{figaro} distribution converges towards the ``final'' distribution, the amount of information that each sample contributes should asymptotically go to zero, hence the information entropy in the reconstructed probability density should go to its limiting value, ultimately determined by the properties, like signal-to-noise ratio, of the signal itself.
We define the entropy as a function of the number of samples $N$ using ~\eqref{entropy} as
\begin{equation}
    S(N) = -\sum_k p_k^{(N)}\log{p_k^{(N)}}\, ,
\end{equation}
where $\{p_k(x)^{(N)}\}_k$ is the distribution obtained with $N$ samples. This quantity is expected to reach a plateau, thus its derivative $\frac{\dd S(N)}{\dd N}$ is expected to go to zero.
Unfortunately, $p_N(x)$ is a realisation from a stochastic process, hence the entropy $S(N)$ and its derivative will exhibit fluctuations. Nevertheless, once the plateau is reached, $\frac{\dd S(N)}{\dd N}$ will randomly fluctuate around zero. Hence, we decided to adopt the number of zero-crossings of $\frac{\dd S(N)}{\dd N}$ rather than some event-specific threshold to decide if the \textsc{figaro} reconstruction converged to its target distribution.
The computation of numerical derivatives of random variables is not trivial either, therefore, we decided to compute $S(N)$ every time a new sample is added to the pool and then perform a linear regression on the values of $S(N)$ that fall within a window of fixed length $L$. The regression gives a ``filtered'' representation of the entropy $S(N)$ in the last $L$ steps. Moreover, we can then approximate $\frac{\dd S(N)}{\dd N}$ the angular coefficient obtained from this linear regression. We found that $L = 500$ is a good choice.

\begin{figure*}
    \begin{subfigure}
    \centering
    \includegraphics[width = 0.6\columnwidth]{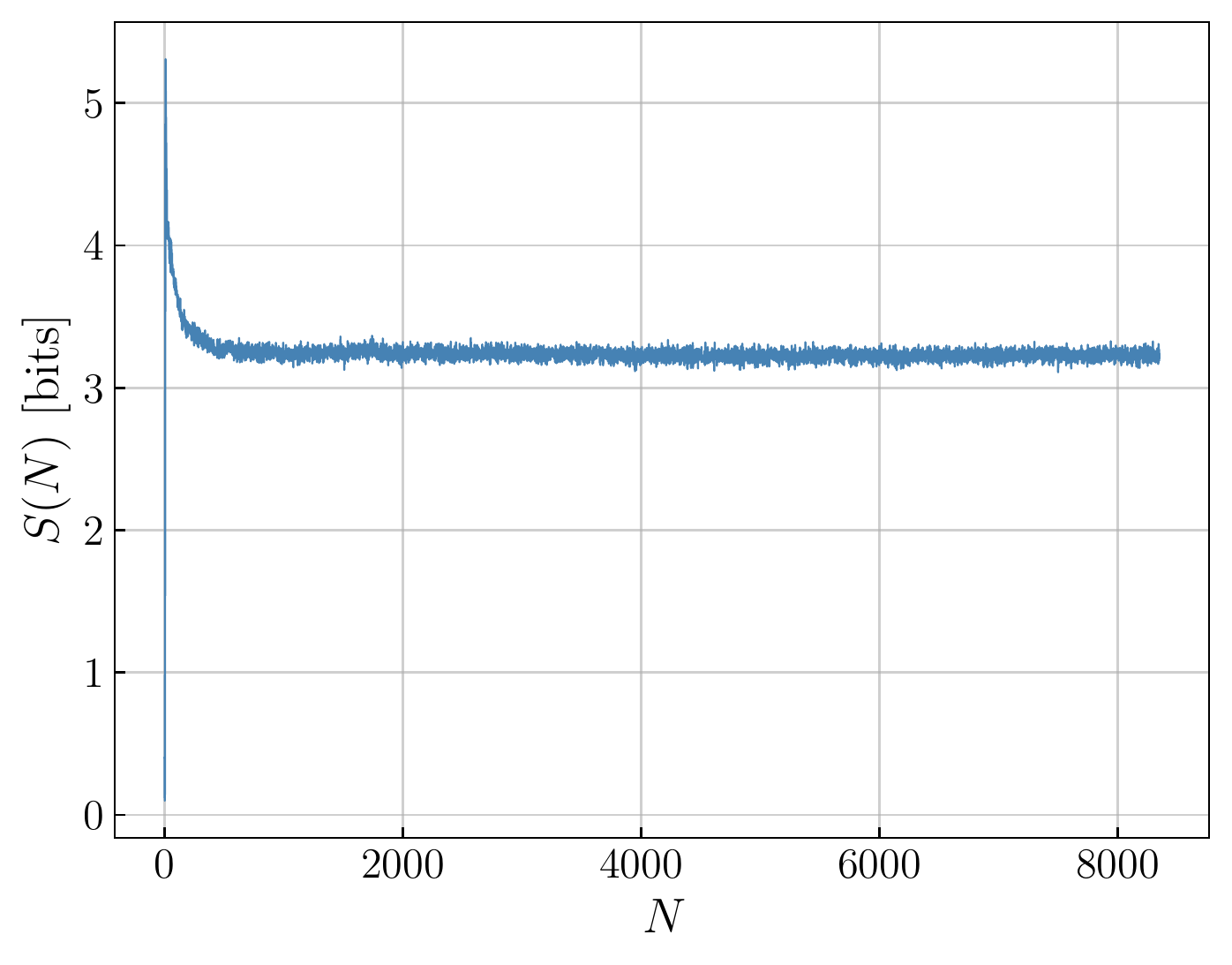}
    \end{subfigure}
    \begin{subfigure}
    \centering
    \includegraphics[width = 0.65\columnwidth]{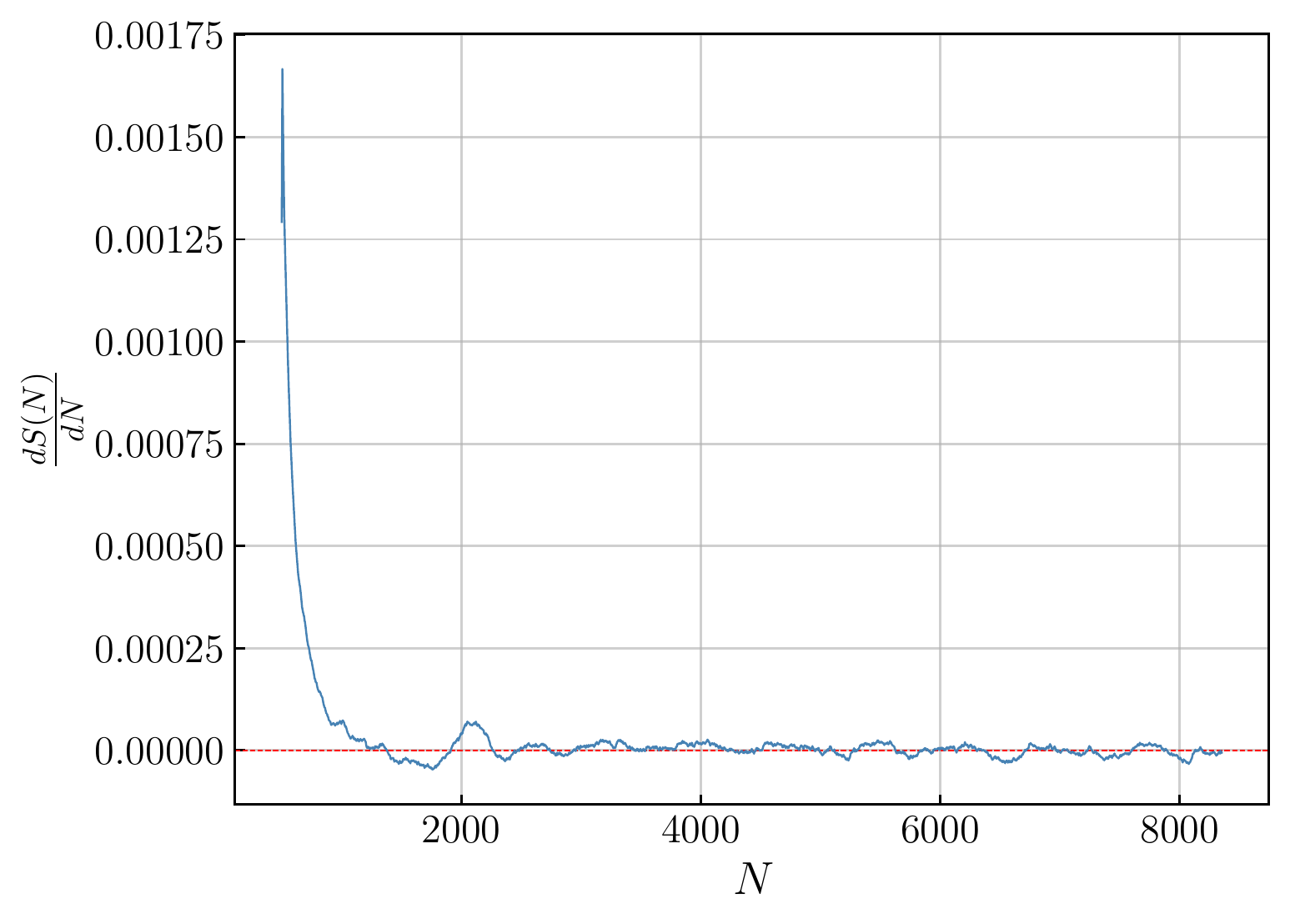}
    \end{subfigure}
    \begin{subfigure}
    \centering
    \includegraphics[width = 0.6\columnwidth]{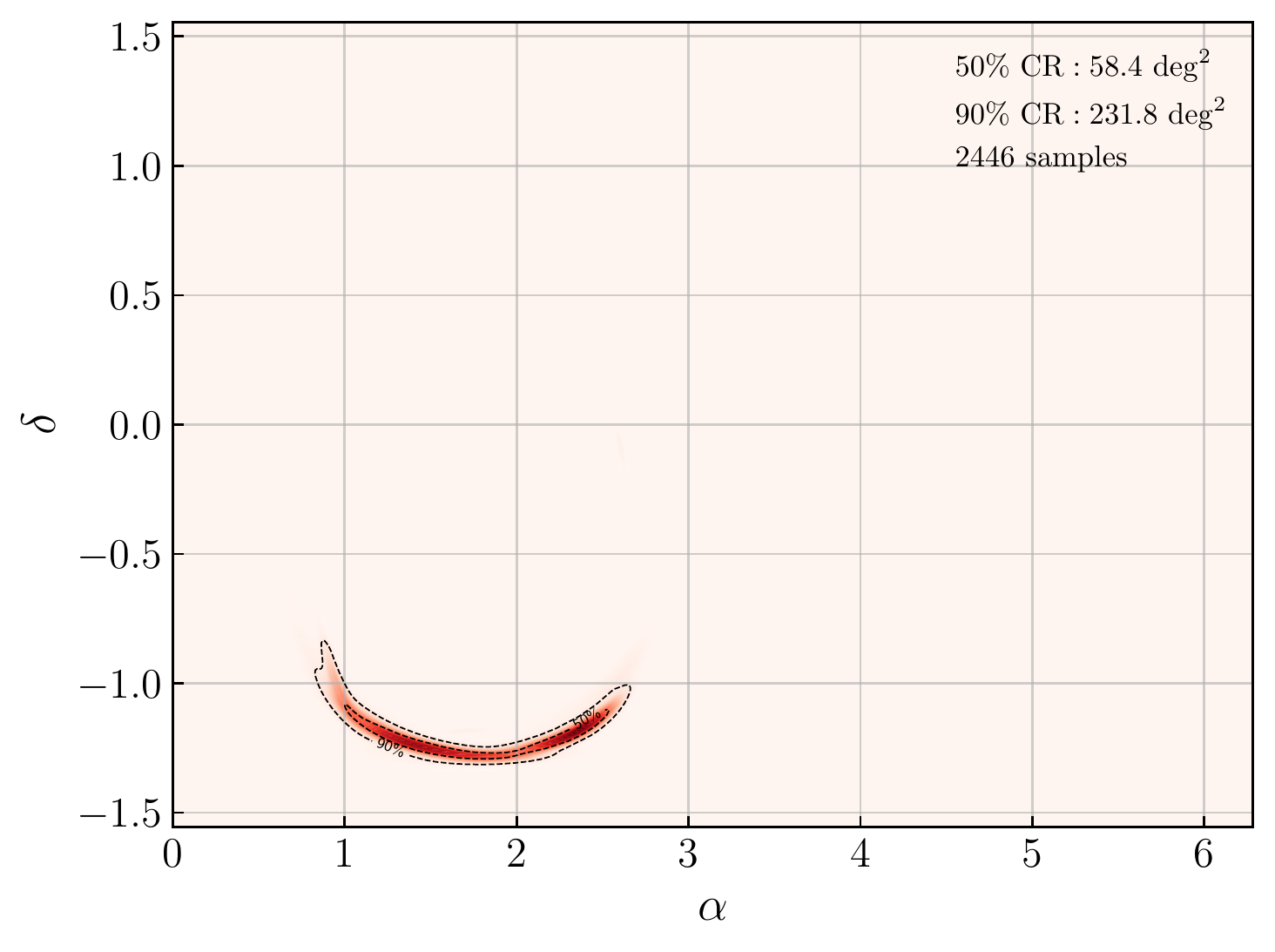}
    \end{subfigure}
    \caption{Entropy (left), entropy derivative (centre) and skymap (right) for GW150914. The skymap is produced using 2446 posterior samples. This number is automatically chosen using the criterion described in Appendix~\ref{app:entropy} with 5 zero-crossings. The window length $L$ is set to 500 samples.\label{fig:entropy}}
\end{figure*}

Figure~\ref{fig:entropy} shows $S(N)$ and $\frac{\dd S(N)}{\dd N}$ for GW150914 along with the skymap produced with the minimum number of samples that satisfied the zero-crossings criterion, 2446 samples. This number corresponds to roughly 30 per cent of all the 8350 available samples. In this case and for illustrative purposes, we set the threshold for producing reliable skymaps at 5 zero-crossings.
The skymap we get is consistent with the one in Figure 2, which is obtained using all the posterior samples.

\section{PP-plots}\label{app:ppplot}
In order to demonstrate the effectiveness of this method, we applied it to the First Two Years of Electromagnetic Follow-up with Advanced LIGO and Virgo, \citep[from now on, F2Y]{f2y} dataset. The simulation campaign consists of thousands of simulated events placed in a stream of coloured Gaussian noise. 250 of these events have been recovered by a real-time detection pipeline and followed up with stochastic samplers to estimate their parameters.

\begin{figure*}
    \begin{subfigure}
    \centering
    \includegraphics[width = \columnwidth]{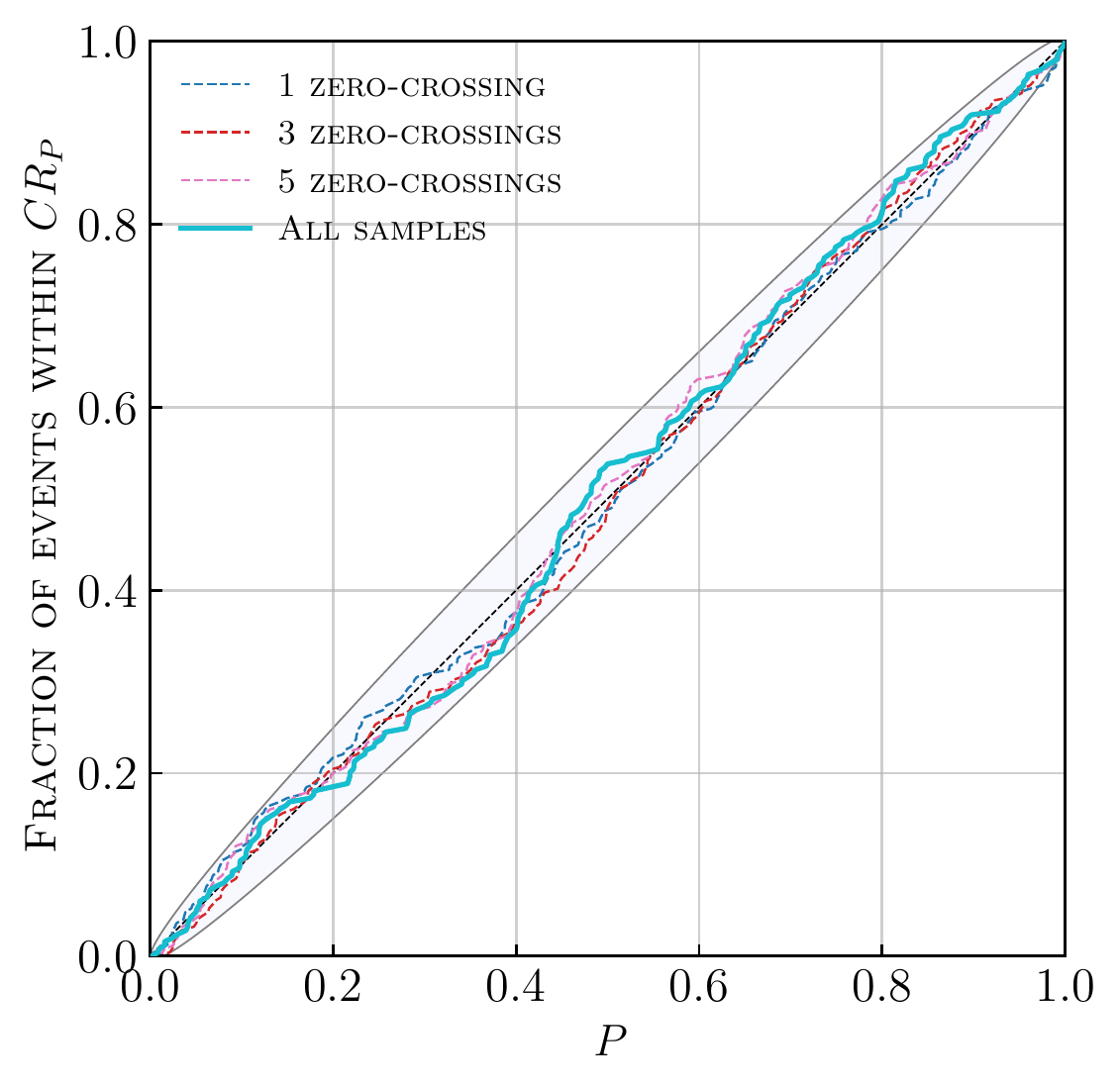}
    \end{subfigure}
    \begin{subfigure}
    \centering
    \includegraphics[width = \columnwidth]{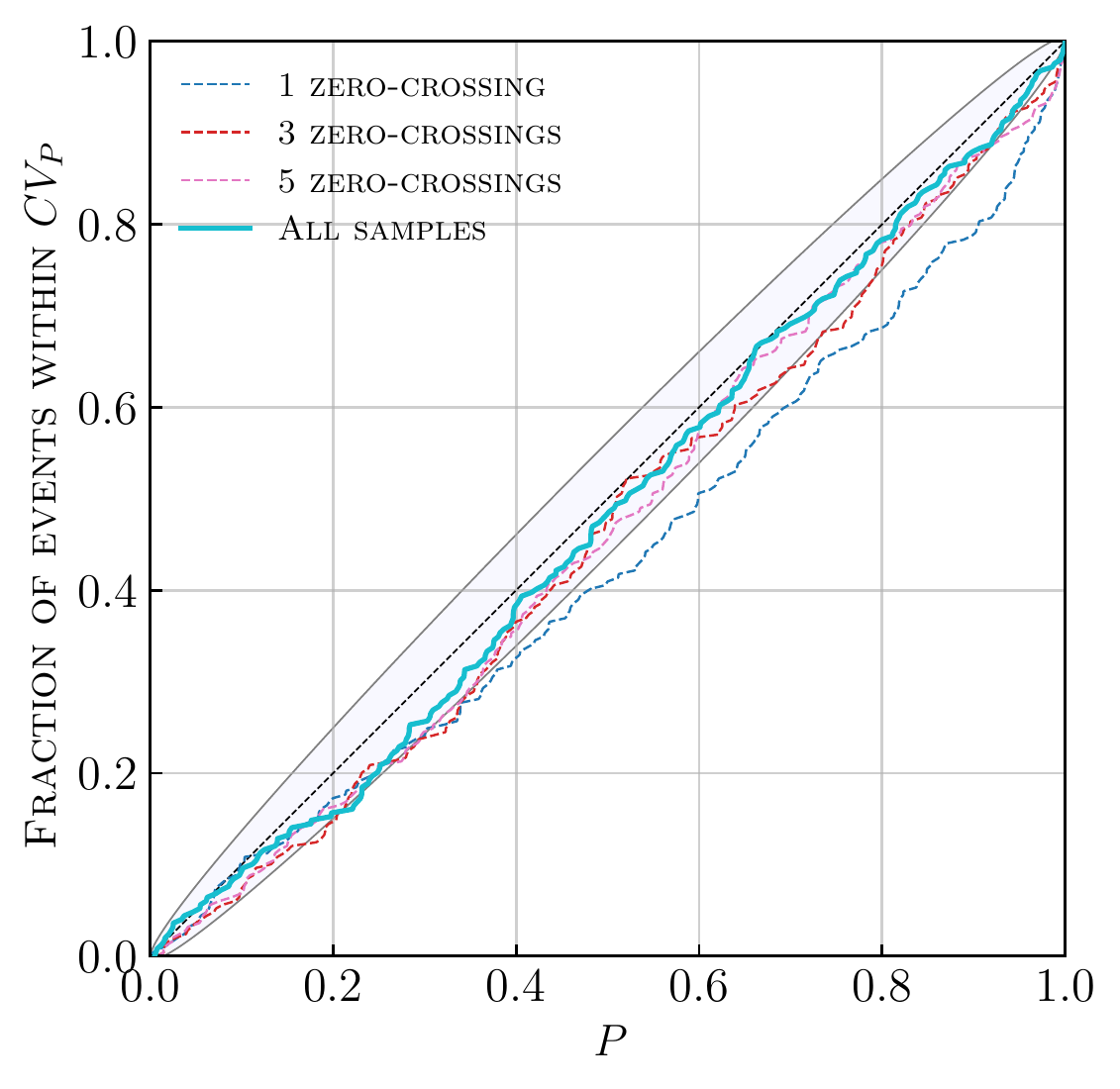}
    \end{subfigure}
\caption{Fraction of events found within a sky-area credible region $CR$ (left) or credible volume $CV$ (right) as a function of encompassed probability $P$. The expected distribution is indicated by the black dashed diagonal line, while the shaded region represents the 95 per cent credible interval for the cumulative distribution, estimated from a beta distribution as in \citet{cameron_2011}. The coloured dashed lines show the cumulative distributions obtained interrupting the posterior reconstruction as soon as the corresponding entropy condition is satisfied for each event, while the thick solid line shows the cumulative distribution obtained using all the available samples for each event.\label{ppplots}}
\end{figure*}

We applied \textsc{figaro} to the posterior samples for the sky position and luminosity distance for all the events available in the F2Y dataset in order to reconstruct the three-dimensional posterior probability density for these events.
In order to verify the self-consistency of the results, we computed the fraction of events located within the credible volume (cfr. region) at a given probability. The expectation from a self-consistent analysis is that a fraction $P$ is found within the $P$-th credible region. Figure~\ref{ppplots} shows the fraction of events found within the credible region $CR_P$ as a function of $P$, both for the 3D volume reconstruction and the sky position alone. We emphasise, however, that since the skymaps are constructed by marginalising the full volume posterior probabilities, the two tests are not independent. In both cases, full volume and sky position, the recovered distribution is in agreement with the theoretical expectations.

Similarly, we tested the entropy-based convergence criterion to establish the threshold necessary to produce reliable skymaps as soon as possible during a MCMC run. We repeated the same exercise of reconstructing the 3D posterior distribution but, instead of using all the available samples, we called the reconstruction ready as soon as the entropy derivative changed sign a certain number of times, see Appendix~\ref{app:entropy}. $\frac{\dd S(N)}{\dd N}$ is computed using a window length of 100 samples.
The pp-plots for 1, 3 and 5 zero-crossings are reported in Figure~\ref{ppplots}.

Focusing on the three-dimensional reconstruction, we see that in this case the recovered distribution is self-consistent for 5 zero-crossings, supporting our proposal of using the number of zero-crossing of the entropy derivative as a indicator of convergence.

\bibliography{bibliography.bib}
\bsp
\label{lastpage}